# Flow Characteristics of Elastically Mounted Slit Cylinder at Sub-critical Reynolds Number

Mayank Verma[1], Alok Mishra[1], Ashoke De[1,2 a)]

[1]*Department of Aerospace Engineering, Indian Institute of Technology Kanpur, Kanpur, 208016, India.*
[2]*Department of Sustainable Energy Engineering, Indian Institute of Technology Kanpur, Kanpur, 208016, India.*

The present work numerically investigates vortex-induced vibrations (VIV) of a two-dimensional circular cylinder with an axisymmetric slit at Reynolds number 500. The study examines the effects of slit shape (i.e., converging, diverging, and parallel slits), the effect of slit-area ratios, slit-width, and its angle with the freestream velocity on aerodynamic forces, vibration response, and associated flow characteristics. The results demonstrate that the addition of the slit assists the VIV suppression by adding an extra amount of flow to the main flow. It results in the stabilized wake with the pressure recovery downstream of the cylinder and causes a reduction in the lift force over the cylinder. Also, there exist different shedding patterns associated with different slit shapes. Proper Orthogonal Decomposition (POD) of the flow field suggests that among all the three slits, the parallel slit heavily modifies the flow behind the cylinder by distributing the amount of energy to a large number of modes as compared to other slits (converging slit and diverging slit), where the most of the energy is contained in a couple of modes. Further, increasing the slit angle (for parallel slit) with respect to the freestream increases the slit effectiveness up to a specific value of slit-angle and beyond that starts to affect the VIV suppression adversely. Observations for the effect of slit-width are also reported from the perspective of suppression of VIV.

## I. Introduction

The dynamics of vortex-shedding behind the bluff bodies have attracted researchers for many years. The interaction of the shedding vortices with the bluff bodies gives rise to the well-known phenomenon of vortex-induced vibrations, VIV. An elastically mounted cylinder, when placed in the fluid flow, starts to oscillate. The amplitude of the oscillations is considerable, especially when the vortex shedding frequency is in synchronization with the oscillation frequency, known as the 'lock-in' region in the literature[1]. This phenomenon can be seen in many engineering fields, viz. buildings, brides, cables, chimneys, etc. VIV has many practical advantages as well as disadvantages, depending on the application. For example, the cables of the long-span bridges are prone to VIV due to their high flexibility and low damping ratios[2]. VIV, in such cases, may cause violent structural vibrations and stresses, leading to considerable fatigue damage. Another example includes VIVACE, Vortex-Induced Vibration Aquatic Clean Energy[3], which is based on the idea of maximizing the vortex shedding and exploiting the VIV for aquatic clean energy conversion. Thus, studies on the different ways to enhance/utilize VIV and suppress VIV have been done experimentally[4-5] and computationally[6-7] in the past. The present study deals with

---

a) Author to whom correspondence should be addressed. Electronic mail: ashoke@iitk.ac.in

the latter one and highlights the role of a slit in controlling the vortex-induced vibrations of an elastically mounted cylinder.

In general, flow control techniques can be broadly categorized as active controls, in which flow control is achieved by supplying the external energy such as the suction and blowing[8], momentum injection[9], forced oscillation of a cylinder at a certain frequency[10], rotating cylinder[11] and the passive controls, where the flow control is obtained without any additional external energy supply. Passive control is obtained by attaching additional devices such as control rods[12], splitter plates[13], or modified surfaces such as grooved or wavy cylinder[14], helical strakes[15]. Both of these control techniques have a significant contribution to the design of bridges, chimneys, buildings. Still, due to external energy supply, the active control methods are complex and expensive from the implementation standpoint.

One of the easiest and simplistic passive methods for flow control is introducing a slit in a cylinder. The use of slit in controlling the vortex shedding dates back to 1978 when Igarashi[16-17] introduced a slit normal to the flow direction in a circular cylinder and observed the stronger and more stable vortex shedding as compared to the unmodified cylinder. Further, Popiel et al.[18] also introduced a concave-rear notch along with the normal slit. They noticed the improved communication of transverse pressure fluctuations between the cylinder's top and bottom surface, leading to stronger and stabilized vortex shedding. Similar observations were also noted by Olsen & Rajagopalan[19] in their experimental study of the cylinder with an axial slit and/or a concave rear notch. Later in 2008, based on the observations from the study by Dong et al.[20] to eliminate the vortex shedding over a circular cylinder using Windward Suction and Leeward Blowing (WSLB), Baek et al.[21] presented a simplified way to implement suction/blowing by placing a slit parallel to the incoming fluid flow. Their study investigated the hydrodynamic effect of small slits on VIV via 2-D and 3-D simulations for different slit widths and observed more than 70% reduction of the Y-amplitude. The parallel slit produces a strong jet flow into the wake and strengthens the secondary vortices attached to the cylinder, resulting in a drastic change in the vortex shedding pattern. In their experimental study of slit cylinders, Junwei et al.[22] observed that the vortex shedding frequency of the cylinder with the slit was faster than the baseline cylinder.

Further, Gao et al.[23-24] experimentally assessed the flow characteristics of the parallel slit placed at a different angle of attacks for $Re = 2.67 \times 10^4$. They observed that for the low angle of attacks, the slit generates a self-issuing jet into the cylinder wake, which reduces the drag and the fluctuating lift component. As the angle of attack approaches 90°, the flow separation point on both sides of the cylinder is delayed, the wake width decreases, and the self-organized boundary layer suction and blowing is observed. Mishra et al.[25], in their initial study on onset of vortex shedding in flow past a circular cylinder with slit, identified two parameters, i.e., slit width ratio (s/D) and slit angle (θ), which plays a crucial role in the determination of critical Reynolds no. ($Re_c$). They concluded that the slit flow increases the pressure downstream the cylinder, resulting in the increased $Re_c$. Another numerical study of viscous flow over the slotted cylinder (with converging, diverging, and parallel slot)[26] reveals that slotted cylinders are although associated with the lesser pressure drag due to extra flow from the slit, exhibit the higher total drag as compared to unmodified cylinder due to the contribution of viscous drag. Recently, Mishra et al.[27] highlighted the role of the parallel slit in the suppression of the vortex shedding and global instability in the laminar regime (Re = 100-500). They observed that between Re = 200 and 300, the primary vortex (cylinder vortex) starts to interact with the secondary vortex (through the slit), and a periodic shedding behavior exists. For Re > 300, the vortex shedding becomes irregular and complex due to the strong interactions between the primary and secondary vortices.

Notably, the available knowledge emphasizes that slit through the cylinder is one of the effective passive flow control techniques for suppressing vortex-induced vibration. All the above studies firmly establish the same. However, most of the studies assumed the cylinder to be stationary/fixed. Very few studies[20-21] considered the actual vortex-induced vibrations of the cylinder. Hence, the motion of the elastically mounted cylinder with slit requires a detailed investigation to understand the flow features and flow physics associated with the cylinder's motion under the VIV. Therefore, the present study attempts to address some pertinent questions: (i) Which slit shape (converging, diverging, or parallel slit) is the most suitable for suppressing the VIV of an elastically mounted cylinder? (ii) In what ways do the slits' area ratio and the slit-width affect the flow downstream of the slit cylinder? (iii) What kind of mode-response exists with the change in the slit angle (for parallel slit) and (iv) what are the wake vortex dynamics associated with the different cases? To address these questions, we have numerically investigated a slit cylinder for three different slit shapes (i.e. converging, diverging, and parallel slits) with parametric variations such as slit-area ratios, slit-width, its angle with the freestream velocity, and their effect on aerodynamic forces, vibration response, and associated flow characteristics.

## II. Problem Statement

The present study examines the VIV suppression over a circular cylinder by using the passive control method of the slit at Reynolds number (based on the cylinder diameter) of 500. The equations are solved numerically over a 2D domain due to the considerable computational time associated with the 3D simulations for such parametric study. The Reynolds number of 500 is chosen based on the observations of Baek et al.[21], where they have reported the insignificant three-dimensional effects on VIV suppression at this Re. However, the authors have also reported the results for the lower value of Re 150 to cover the laminar flow regime. The study emphasizes the fluid dynamics involved in suppressing vortex shedding for different slit shapes (i.e. converging, diverging and parallel slit), resulting in the reduced VIV. The shapes are examined for various slit-area ratios and at different angles, w.r.t. the freestream flow. The circular cylinder is elastically mounted with the help of spring and restricted to vibrate in transverse/crossflow direction only. Although the present study deals with the motion of a physically unconfined rigid cylinder of diameter D, the numerical discretization requires finite boundaries in the far-field. The typical 2-D computational domain varies between $-20 \leq x/D \leq 30$ and $-20 \leq y/D \leq 20$, considering the center of the cylinder located at the origin. The incoming flow is assumed to be steady and uniform, while the zero Neumann boundary condition is used for flow velocity at the outlet. Although, the authors have also checked for continuative boundary condition[28], i.e. vanishing the second derivative at the outlet for one case similar to Biswas and Chattopadhyay[29], and found no significant variation in the VIV characteristics of the cylinder. Hence, the zero Neumann boundary condition is used at the outlet for this work. The top and bottom sides of the domain are treated with the freestream boundary condition and the no-slip boundary condition at the cylinder wall.

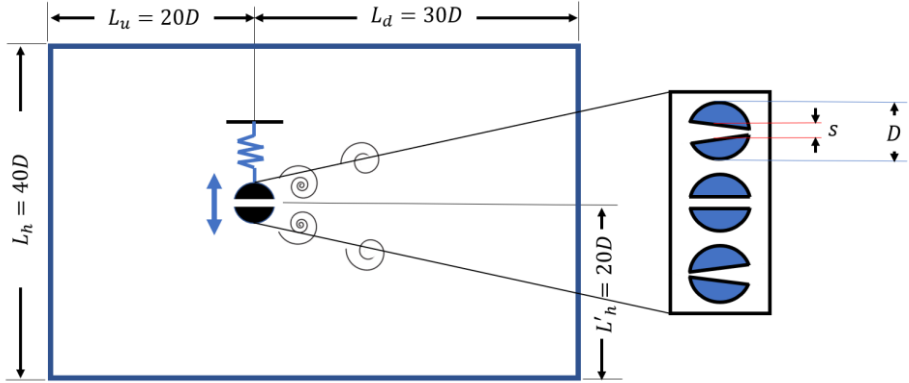

FIG. 1. Schematic of the computational domain for the study.

## III. Numerical Setup

### A. Flow Solver

The flow is assumed to be incompressible, laminar, and viscous. To model the fluid flow, we solve the conservation equations for mass and momentum as given follows,

$$\nabla \cdot \vec{v} = 0 \tag{1}$$

$$\rho_f \left[ \frac{\partial \vec{v}}{\partial t} + (\vec{v} \cdot \nabla)\vec{v} \right] = -\nabla p + \mu_f \nabla^2 \vec{v} \tag{2}$$

Where $\vec{v}$ represents the velocity vector of the fluid, $\rho_f$ is the fluid density, p is the static pressure, and $\mu_f$ is the dynamic viscosity of the fluid. The Reynolds number based on the inlet velocity and the cylinder diameter is defined as, $\text{Re} = \dfrac{UD}{\upsilon}$, where $\upsilon = \mu_f / \rho_f$, is the kinematic viscosity of the fluid. The study is performed at two Reynolds numbers, Re = 150 and Re = 500, to assess the behavior of the slit in suppressing the vibrations. Open-source CFD solver OpenFOAM[30] is used to solve the flow conservation equations. PIMPLE algorithm is used to address the pressure-velocity coupling. Second-order discretization schemes are used for all the spatial and temporal terms in the governing equations.

To incorporate the mesh motion due to movement of the cylinder, the mesh velocity field has been computed by Laplace's equation,

$$\nabla \cdot (\gamma_m \nabla z) = 0 \qquad (3)$$

Where $\gamma_m$ is the diffusion coefficient, and $z$ is the mesh displacement field. Due to the fixed top and bottom boundaries, the mesh motion is distributed through the grid using the inverse mesh diffusion model, which calculates the mesh diffusion based on the inverse distance from the cylinder body[31-32]. It ensures that the farther away from the specified moving body, the less mesh morphing. The diffusivity field is based on the quadratic relation on the inverse of the cell center distance to the nearest boundary: ($1/l^2$).

## B. Structural Solver

The cylinder is constrained to allow movement only in the cross-flow direction. The equation of motion for a 1-DOF single-cylinder motion is assumed to be linear in the transverse direction only without any rotation and can be written as,

$$m\dfrac{d^2Y}{dt^2} + 2m\zeta\omega_n \dfrac{dY}{dt} + kY = F_Y \qquad (4)$$

Where *m* is the mass of the system (including the cylinder mass and the added mass), $\omega_n$, $\zeta$ and k are the natural frequency of the system $\left(\omega_n = \sqrt{\dfrac{k}{m}}\right)$, damping coefficient $\left(\zeta = \dfrac{c}{2\sqrt{km}}\right)$, and the spring constant, respectively. The external fluid forces are obtained from the flow solver as external forces (pressure + viscous forces), and the corresponding displacement value, Y, and the structural velocity, $\dot{Y}$, is calculated simultaneously and fed to the mesh motion module to shift the cylinder to the updated position. The transverse motion is updated using the second-order Runge-Kutta Mid-point Scheme. The motion of such vibrating systems can be described effectively using some of the non-dimensional parameters such as non-dimensional mass-ratio $\left( m^* = \dfrac{m}{\dfrac{\pi}{4}\rho D^2 L} \right)$, non-dimensional spring constant $\left( k^* = \dfrac{k}{\rho U_\infty^2 L} \right)$, non-dimensional damping coefficient $\left( C^* = \dfrac{c}{\rho U_\infty D L} \right)$, non-dimensional velocity ratio $\left( U_r = \dfrac{U_\infty}{f_n D} \right)$, non-dimensional frequency ratio $\left( f = \dfrac{f_s D}{U} \right)$, non-dimensional time $\left( \tau = \dfrac{tU}{D} \right)$, non-dimensional oscillation amplitude ($Y/D$), and non-dimensional mean oscillation amplitude $\left( A^* = \dfrac{(Y/D)_{max} + abs(Y/D)_{min}}{2} \right)$.

## C. Proper Orthogonal Decomposition (POD)

The proper orthogonal decomposition (POD) is one of the classical ways to decompose the flow field in time and space. Kosambi[33] first proposed this and later extended it by other researchers like Lumley[34] and Sirovich[35], who was the first one to propose a computationally inexpensive "Method of Snapshot". POD is a powerful tool to extract dominant flow structures based on their energy contribution. The computations of eigenmodes are carried out by collecting a sufficient number of snapshots to maintain the temporal resolution of the decomposition process. Evidently, the decomposition process confirms that only a few basis functions can represent most of the energy; only the first few modes are sufficient to represent the flow field accurately. Previous works of the present group extensively used this technique for different flow configurations[36-42].

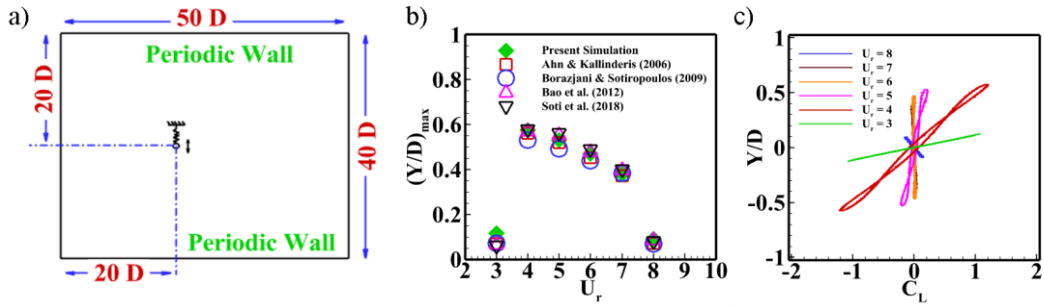

FIG. 2. (a) Schematic of the benchmark VIV used to validate the computational framework, (b) comparison of maximum displacement with the published results, and (c) Phase portraits of the transverse force relative to the transverse motion for the 1-DOF motion of the unconfined cylinder

## D. Numerical Validation

Before addressing the main problem statement, we have performed the numerical validation study for the 1-DOF VIV system with a circular cylinder mounted on the elastic supports employing spring. Figure 2 (a) shows the computational domain used for the validation study. The mass ratio of the cylinder is 2, the damping coefficient is 0, and the Reynolds number is kept fixed at 150.

In Fig. 2 (b), the results of our numerical setup are compared with the published results available in the literature and are found to be in good agreement. The maximum value of the non-dimensional amplitude $(Y/D)_{max}$ is found to be 0.5742 in our simulations. The same was predicted to be 0.5624, 0.5312, 0.5680, and 0.5782 by Ahn and Kallinderis[43], Borazjani and Sotiropoulos[44], Bao et al.[45], and Soti et al.[46] and differs by ~2 % from our results. Figure 2 (c) represents the phase portraits of the VIV motion for different velocity ratios, which further confirms the lock-in region to be spanned over $3 \leq U_r \leq 7$.

## E. Grid Independence and Error Analysis

As the study deals majorly with the cylinder with slit, the grid independence study is performed over the transversely vibrating cylinder with a parallel slit of slit-width (s/D) as 0.20 at Re 150. ANSYS ICEM-CFD[47] is used to generate mesh over the domain. O-grid blocking is used in the near cylinder region to map the cylinder surfaces, as portrayed in Fig. 3 correctly. The y+ value at the cylinder surface and the slit is kept below 1 to resolve the wall stresses correctly.

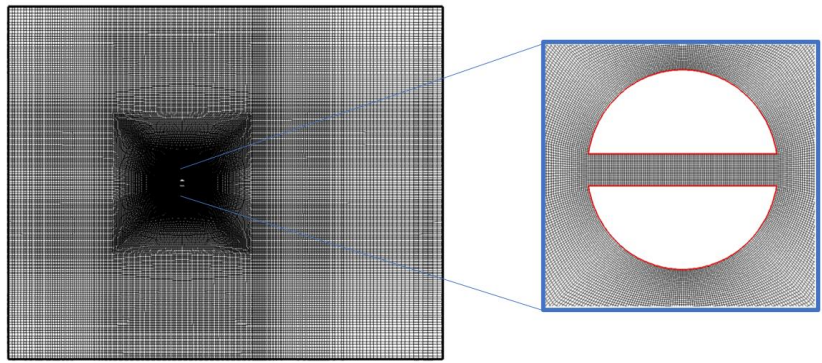

FIG. 3. Computational grid over the circular domain

The mass ratio is kept at 2, while the velocity ratio and damping ratio are considered 4.44 and 0, respectively. The grid independence study is conducted on four sets of grids i.e., Grid-A [37,217 cells], Grid-B [59,665 cells],

Grid-C [93,976 cells], and Grid-D [147,192 cells]. The variation of the aerodynamic load coefficients, i.e., $C_L$ over the cylinder and the rms value, is plotted in Fig. 4 (a) and (b).

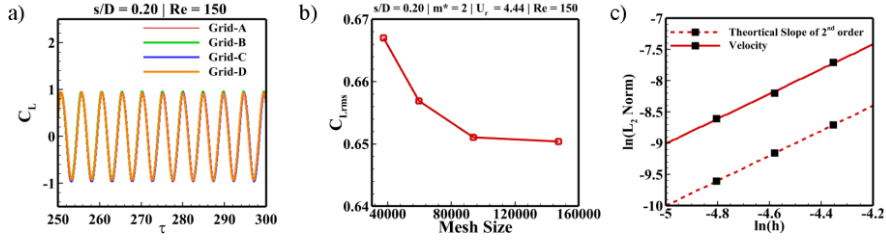

FIG. 4. Grid Independence Study at Re = 150 for an unconfined cylinder with a parallel slit (a) $C_L$ variation with time, (b) $C_{L\,rms}$ variation with the mesh size, (c) $L_2$ the norm against the grid spacing h.

As observed in Fig. 4 and Table I, there is no significant difference in Grid-C and Grid-D results. Hence, it is concluded that the solution becomes grid-independent at Grid-C and is selected for further simulations.

TABLE I Grid Independence study for the moving cylinder

|  | No. of elements | $C_{L\,rms}$ | % Change in $C_{L\,rms}$ value |
|---|---|---|---|
| Grid-A | 37,217 | 0.6657 | --- |
| Grid-B | 59,665 | 0.6526 | 1.96 % |
| **Grid-C** | **93,976** | **0.6502** | **0.36 %** |
| Grid-D | 147,192 | 0.6490 | 0.18 % |

To develop more confidence in the grid, we have also performed the grid-convergence study using the Grid Convergence Index (GCI) proposed by Roache[48-49]. A detailed description of the GCI is given in Appendix.

Table II reports the GCI for rms value of the lift coefficient ($C_{L_{rms}}$) for three different grids (Grid-B, Grid-C, and Grid-D). For the best estimation of the grid convergence relating to a 50% grid refinement, the Factor of safety ($F_s$) is taken as 1.25[50]. The value of GCI reduces for the constitutive grid refinements for both the parameters. The GCI calculations confirm that the Grid-C is nicely resolved.

TABLE II. Richardson error estimation and grid-convergence index for three sets of grids

|  | $r_{CB}$ | $r_{DC}$ | o | $\varepsilon_{CB}$ ($10^{-2}$) | $\varepsilon_{DC}$ ($10^{-3}$) | R | $E_2^{coarse}$ | $E_1^{fine}$ | $GCI^{coarse}$ | $GCI^{fine}$ |
|---|---|---|---|---|---|---|---|---|---|---|
| $C_{L\,rms}$ | 1.25 | 1.25 | 1.927 | -3.7759 | -1.8455 | 0.49 | 0.0109 | 0.00223 | 1.36 % | 0.28 % |

## IV. Results and Discussion

We have used the flow visualization technique and the Proper Orthogonal Decomposition (POD) technique to capture and understand the fluid dynamic phenomena associated with the different slit shapes (i.e., converging slit, diverging slit, and parallel slit). The vorticity contours, streamline patterns, and the aerodynamic lift force for different shapes are compared to the cylinder with no slit case for stationary and moving cases. The frequency spectrum of the lift force coefficient and the transverse amplitude of the oscillation is plotted to identify the dominant frequencies that affect the flow physics. Further, fixed slit width s/D = 0.16, based on Baek et al.[21], is identified for the numerical investigations. The effect of Reynolds number, slit area ratio, slit width on different slit shapes, and the slit angle (for parallel slit only) is investigated to assess the VIV suppression. Four different area ratios (maintaining the slit height, s/D=0.16, constant at the center of the cylinder) 1.5, 2.3, 3, and 4 are reported in the paper. Further, the angle between the freestream flow and the slit (i.e., slit angle, $\theta$) is varied from 0° to 30° with a step increment of 5° for parallel slit case.

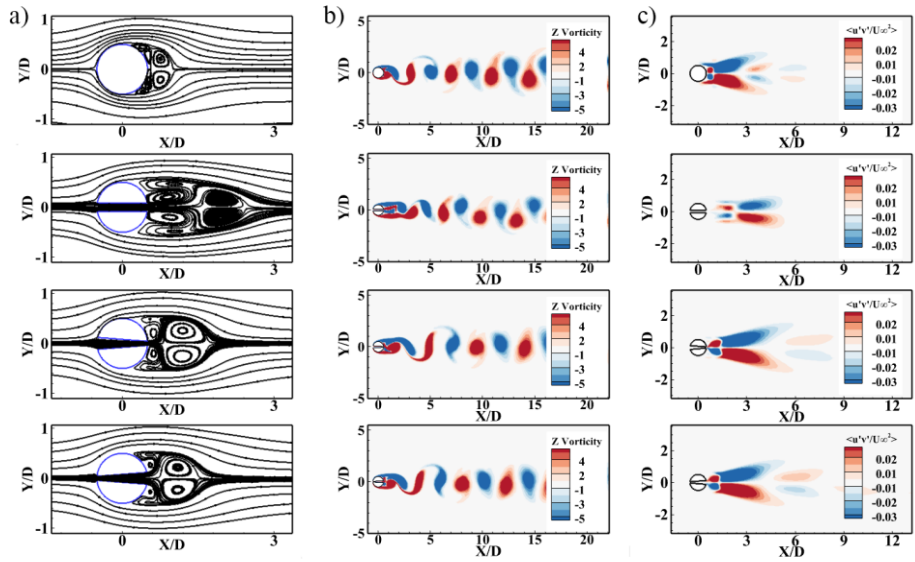

FIG. 5. (a) Streamlines drawn from the time-averaged velocity field for stationary cylinder; (b) Instantaneous vorticity contour plots for different slits compared with the no slit cylinder case; and (c) Normalized Reynolds Shear Stress for different slits. [at Re = 500, slit width (at center)= 0.16, and slit-area ratio = 3]

## A.  Effect of Slit Shape

### 1.  Stationary Cases

Firstly, we investigate the effect of slit shapes for three different shapes (i.e., converging slit, parallel slit, and diverging slit) using 2-D simulations with a stationary cylinder under uniform flow conditions at Re = 500. The slit height (calculated at the center of the cylinder) is fixed at s/D = 0.16, and the slit-area ratio (the ratio between slit height at the entry and the exit of the slit for the converging slit and vice versa for the diverging slit) is fixed at 3 (for converging and diverging slit) and 1 (for parallel slit). Fig. 5 (a) and (b) portray the time-averaged velocity streamlines and the instantaneous vorticity contour for the three slit shapes. For the stationary cylinder case, all the slit shapes shed vortices in the same shedding pattern as the reference case of the cylinder with no slit. Although, the additional flow through the slit creates the secondary vortices, which further interact with the primary vortices shed from the cylinder surface. Also, the extra momentum of the flow through the slit shifts the vortex formation to the downstream position, as shown in Fig. 5 (a).

Figure 5 (c) provides the Reynolds Shear stress field behind the cylinder, calculated from the mean of the product of the fluctuating velocity components normalized with the square of the freestream velocity $\left(u'v'/U_\infty^2\right)$. The calculation of the fluctuating component of the velocity involves the difference between the instantaneous velocity and the mean velocity at that instant. These statistics are collected after 120 flow-through times when the flow characteristics show a quasi-steady periodic variation. The distribution of the Reynolds shear stress is anti-symmetric about the wake-axis for all the cases. Although, the slit flow results in the lower peak values for the Reynolds shear stress (lowest being for the parallel slit case). This represents the weakening of the vortex shedding and supports the wake stabilization due to slit. The peak values of the Reynolds shear stress are also shifted to the downstream location away from the cylinder base.

Further, Fig. 6 draws the non-dimensional mean axial velocity at the center-line. This shows that the flow through the slit translates the vortex bubble to the downstream location and elongates it in the axial direction. The length of the recirculation bubble is calculated based on the distance between the points where the mean axial velocity changes its sign. One can observe the maximum shift of vortex pattern for the case of the parallel slit. At the same time, the maximum elongation of the bubble is observed for the case of the convergent slit.

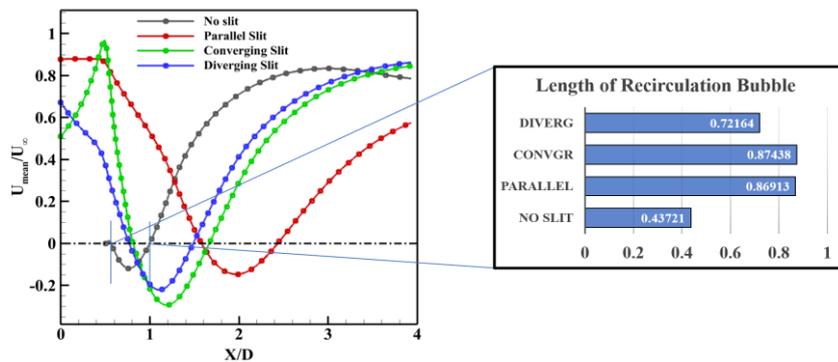

FIG. 6. Distribution of Mean axial velocity at the centerline for different slits and their corresponding length of the recirculation bubble

To further investigate the cause of these shifts, we look at the velocity development inside the slit and in the region behind the cylinder at different axial locations shown in Fig. 7. As expected, owing to its shape, the divergent slit reduces the velocity of flow inside the slit, the convergent slit increases the velocity, while for the parallel slit, the velocity is almost constant at the center and throughout the slit (can be seen in Fig. 7(a) and (b)).

As observed, the momentum of the fluid coming out of the slit is maximum for the parallel slit case (as the area under the velocity curve is maximum in Fig. 7(c) and (d)). This higher momentum for the case of the parallel slit causes the maximum shift to the vortex formation, resulting in the pressure increase in the region downstream of the cylinder. This pressure increase is visible in the Mean coefficient of Pressure ($C_{p\,Mean}$) distribution behind and over the cylinder surface in Fig. 8.

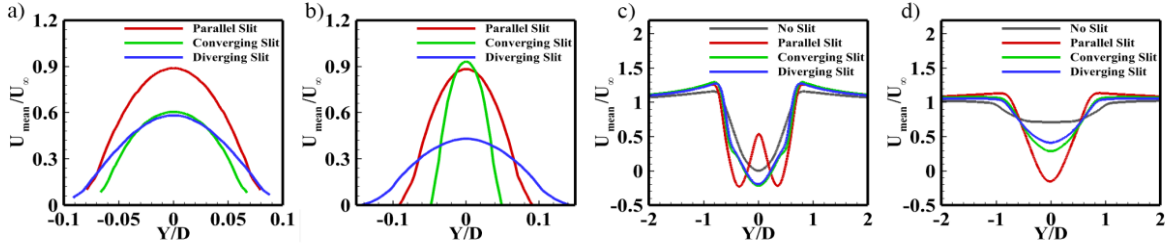

FIG. 7. Distribution of the time-averaged axial velocity inside the slit at different axial locations (a) x/D = 0.15, and (b) x/D = 0.45 (corresponding to the slit exit), and the time-averaged axial velocity distribution behind the cylinder (c) x/D = 1, and (d) x/D = 2.

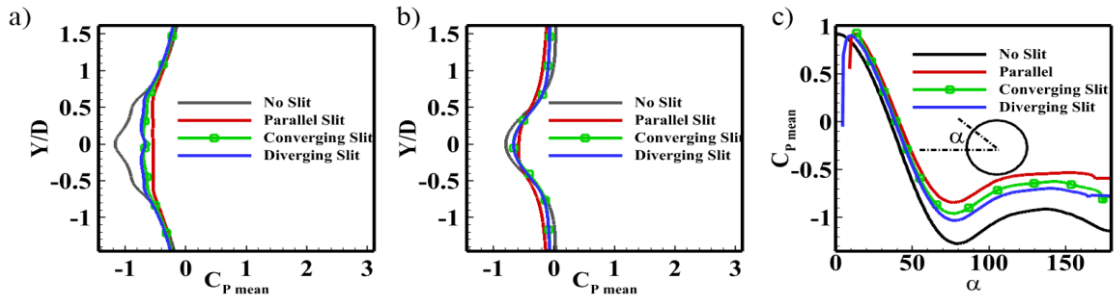

FIG. 8. Distribution of mean pressure coefficient ($C_{p\,Mean}$) behind the cylinder (a) at x/D = 0.6, (b) at x/D=1.0; and (c) over the stationary cylinder surface for different slits

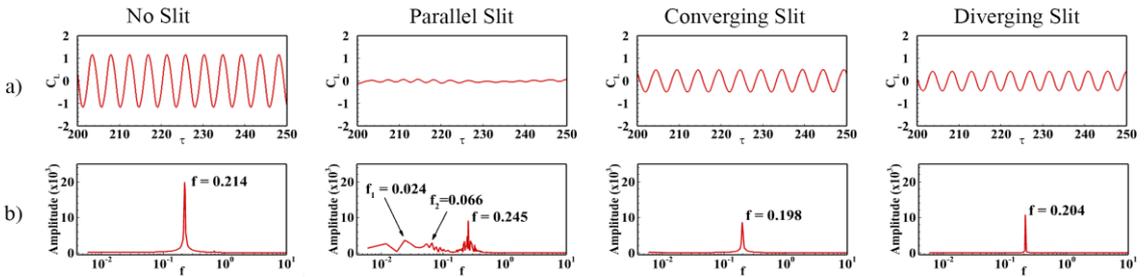

FIG. 9. (a) Temporal distribution of lift coefficient ($C_L$) over the cylinder surface and (b) their corresponding Fast Fourier Transform (FFT).

Further, Fig. 9 compares the aerodynamic loading coefficient (i.e., Lift force-coefficient, $C_L$) for all the different cases and presents the lift coefficient against the non-dimensional time, $\tau$, and the frequency distribution of the Lift-coefficient. Introducing the slit in the cylinder reduces the lift force on the cylinder for all the slit shapes due to the weakening of the vortex shedding. Owing to the higher momentum fluid from the slit, the Parallel slit performs better than the converging and diverging slit for the reduction of Lift force. Also, the frequency spectrum of the lift-coefficient yields that the converging and diverging slit retains the super-harmonics nature of frequency distribution with clearly identifiable super-harmonics (f = 0.198 for converging slit and f = 0.204 for diverging slit) similar to no slit case (f = 0.214), while the parallel slit shows the frequency distribution drastically different from the other cases. The peaks in the no-slit case completely disappear, and no. of peaks at much lower frequency dominate. Baek et al.[21] termed it the *detuning process* caused by the parallel slit.

All the above discussions support that the parallel slit is associated with the lowest aerodynamic loading for the stationary cases. As the lift force is the driving force for the VIV, it might be possible that the cylinder with the parallel slit exhibits the lowest amplitude of vibrations when subjected to the fluid flow over it. Further, to support our argument and investigate the various flow features associated with the oscillating cylinder, the VIV cases with the oscillating cylinder are presented in the coming section.

## 2. VIV Cases

Simulations with the stationary slit cylinders develop a basic understanding of the effectiveness of the slit shapes. Further, to confirm the effectiveness of the slit shapes in the moving cases, the simulations with the 1-DOF transverse motion are carried out with and without different slits. The velocity ratio ($U_r$) is 4.44, corresponding to the maximum displacement amplitude, as shown in Fig. 2. The damping coefficient is set to 0 to maximize the transverse motion. The mass ratio is set to 2 as the larger mass ratios are associated with smaller amplitudes (Baek et al.[21]).

Figure 10 illustrates the flow distribution behind the oscillating cylinder via means of instantaneous vorticity magnitude, Reynolds shear stress, and their time histories of the transverse amplitude response (Y/D) of the cylinders as well as the corresponding lift coefficient ($C_L$). The results identify two vortex shedding modes for different slit shapes (2S mode and S+P mode) classified by Williamson[51]. The cylinder with no slit vibrates under the influence of VIV, and the vortex shedding pattern resembles the S+P mode of the vortex shedding, which corresponds to the one single vortex with a pair of counter-rotating vortices in each cycle. With the parallel slit, the vortex shedding is chaotic and resembles the 2S mode. The converging slit exhibits the S+P mode of the vortex shedding, while the diverging slit sheds the alternate vortices and forms a well-known Karman vortex street resembling the 2S mode of the vortex shedding.

Figure 10 (b) depicts the Reynolds shear stress distribution in the cylinder wake. The distribution is anti-symmetric about the wake axis. Also, the reduced Reynolds shear stress values reduce behind the cylinder with the slit. The parallel slit shows the lowest peak values for the Reynolds shear stress, and the peak values are shifted away from the base of the cylinder. This results in the reduced velocity fluctuations in the wake and the reduced lift coefficient values over the cylinder surface. Fig. 10 (c) yields that for the cases with the slit, there is a significant reduction in the transverse amplitude compared to the no slit case. Figures 11 (a) and (b) report the pressure distribution downstream the cylinder at x/D = 0.6 and at the cylinder surface for different slits, respectively. There is a maximum pressure recovery in the case of the cylinder with the parallel slit. Also, the phase portraits drawn in Fig. 11 (b) suggest that the cylinder with no slit oscillates with out-of-phase lift force, while the cylinder with slits oscillates with the same phase as the lift force. Figure 11 (c) compares the maximum mean amplitude ($A^*$) of the transverse motion for all the slits and the no slit case to felicitate the better comparison at two different Reynolds numbers, i.e., Re=150 and Re=500. The chart establishes the role of the parallel slit in reducing the VIV, making it the most suitable slit shape for the VIV suppression among the three. Also, with the increase in the Reynolds number, the effectiveness of the parallel slit increases.

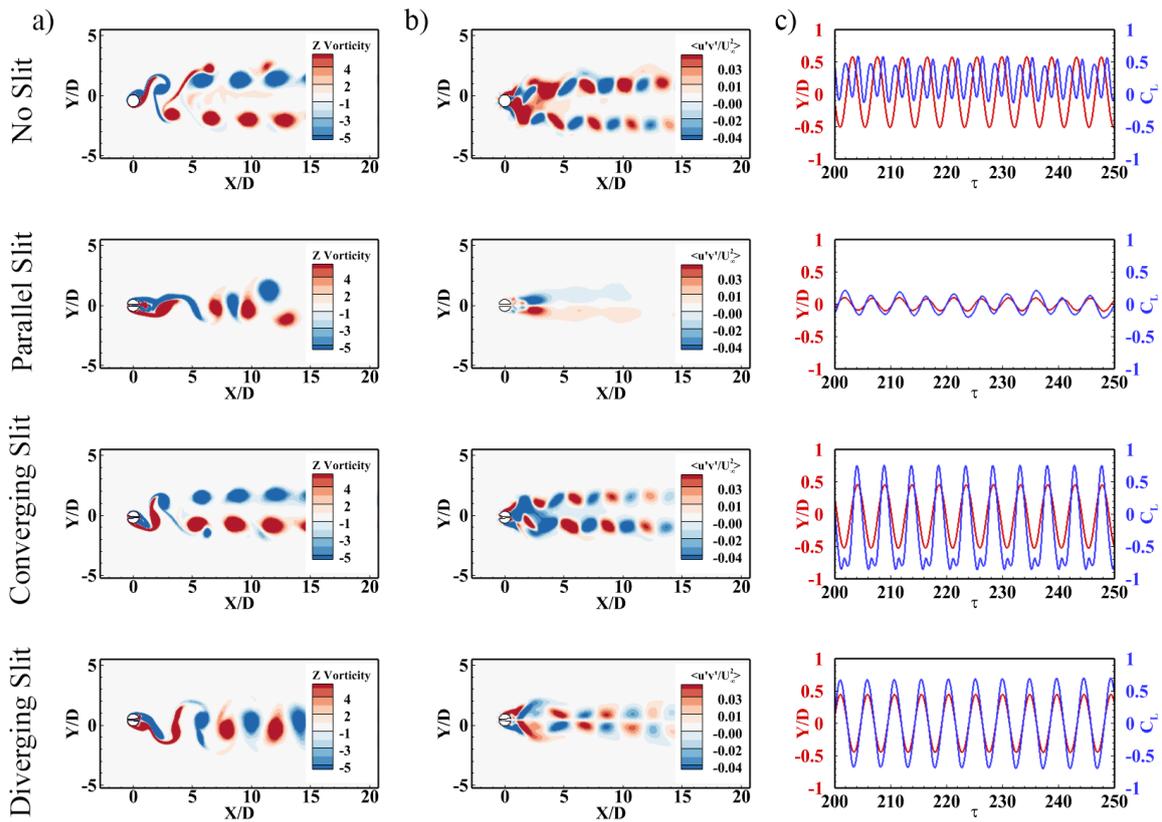

FIG. 10. (a) Instantaneous vorticity contour (b) Normalized Reynolds Shear Stress (c) Coefficient of the transverse lift force ($C_L$) and corresponding transverse displacement of the cylinder (y/D), [at Re = 500, $U_r = 4.44$, and $m^* = 2$, $C^* = 0$, s/D = 0.16 and slit-area-ratio = 3]

Following the global instability representation of the flow oscillation in terms of the maximum value of rms fluctuation intensities over the whole domain as suggested by Delaunay and Kaiksis[52], Figs. 12 (a) and (b) report the maximum values of rms fluctuation intensities ($u'_{rms}$ and $v'_{rms}$) by extracting the values in different locations of x/D. The interesting thing to observe is the reduced peak values of the maximum fluctuating velocity component for the slit cylinder cases. This supports our above statement of stabilizing the flow through reduced Reynolds shear stress values for the cases with the slit. Further, Nishoka and Sato[53] reported that the peak values of the fluctuating velocities shift upstream (close to the cylinder) with an increase in the Reynolds number for the flow over a normal cylinder. Following this, one can say that the cylinder with the parallel slit experiences a reduced Reynolds number flow as the peak values shift downstream for the parallel slit (away from the cylinder) compared to the no-slit case. As a result, the lift coefficient over the cylinder surface reduces and provides a maximum suppression of the VIV when a parallel slit is introduced.

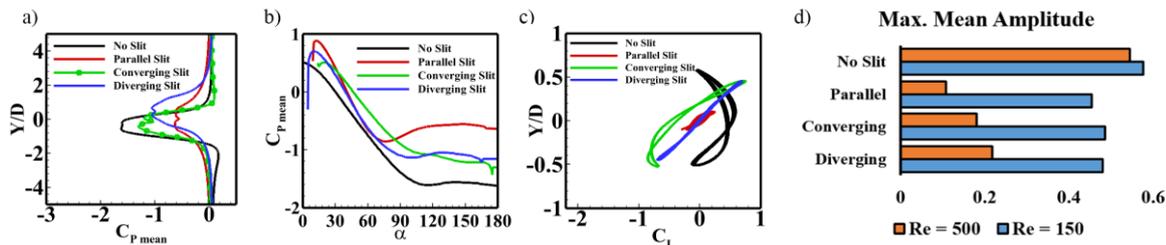

FIG. 11. Distribution of Coefficient of Pressure (a) behind the cylinder at x/D = 0.6; (b) over the cylinder surface for different slits; (c) Phase Portraits for the transverse force relative to the transverse motion; (d) Maximum mean transverse amplitude of the transverse oscillation of the cylinder at two different Reynolds number (Re = 150 [Blue] and Re = 500 [Orange]).

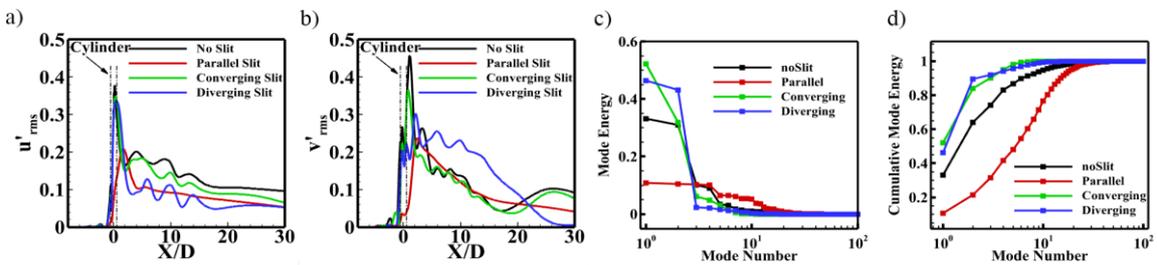

FIG. 12. The maximum rms velocity fluctuations along the line of constant x/D for different slits: (a) Axial velocity fluctuations (u'), (b) Transverse velocity fluctuations (v'); Distribution of energy over POD modes: c) Mode energy; (d) Cumulative mode energy.

Table III reports the first ten eigenvalues for the demonstration of energy distribution over POD Modes. Further, to understand the slit flow interaction with the primary wake and its associated structures, the decomposition of the flow field in the number of energy modes is felicitated with the help of POD. POD, a mathematical tool to evaluate the dominant flow structure based on its energy contribution, provides spatial coherence amongst the structures by extracting the spatial orthogonality. A table of POD modes based on the fluctuation kinetic energy can be derived from the snapshots of the instantaneous flow field, where the extracted

POD modes are proportional to their eigenvalues. A snapshot independence study for the POD modes with 50, 100, and 150 snapshots has been performed, and no significant difference is observed in the energy modes for the 100 and 150 snapshots. Hence, 100 snapshots are found to be sufficient enough for the present study. Figures 12 (c) and (d) illustrate the energy distribution over the 100 POD modes for various slit shapes.

TABLE III. First ten eigenvalues for the demonstration of energy distribution over POD Modes

| Mode | No Slit | Parallel Slit | Converging Slit | Diverging Slit |
|---|---|---|---|---|
| I | **33.16** | **10.83** | **52.17** | **46.42** |
| II | **30.93** | **10.50** | **31.85** | **43.14** |
| III | **10.14** | **10.21** | **6.18** | **2.32** |
| IV | **8.94** | **10.08** | **4.77** | **2.08** |
| V | **3.40** | **6.41** | **2.94** | 1.52 |
| VI | **3.08** | **6.38** | 1.33 | 1.21 |
| VII | 1.70 | **6.04** | 0.43 | 1.122 |
| VIII | 1.31 | **5.44** | 0.15 | 0.59 |
| IX | 1.08 | **5.37** | 0.06 | 0.56 |
| X | 0.98 | **5.19** | 0.03 | 0.25 |

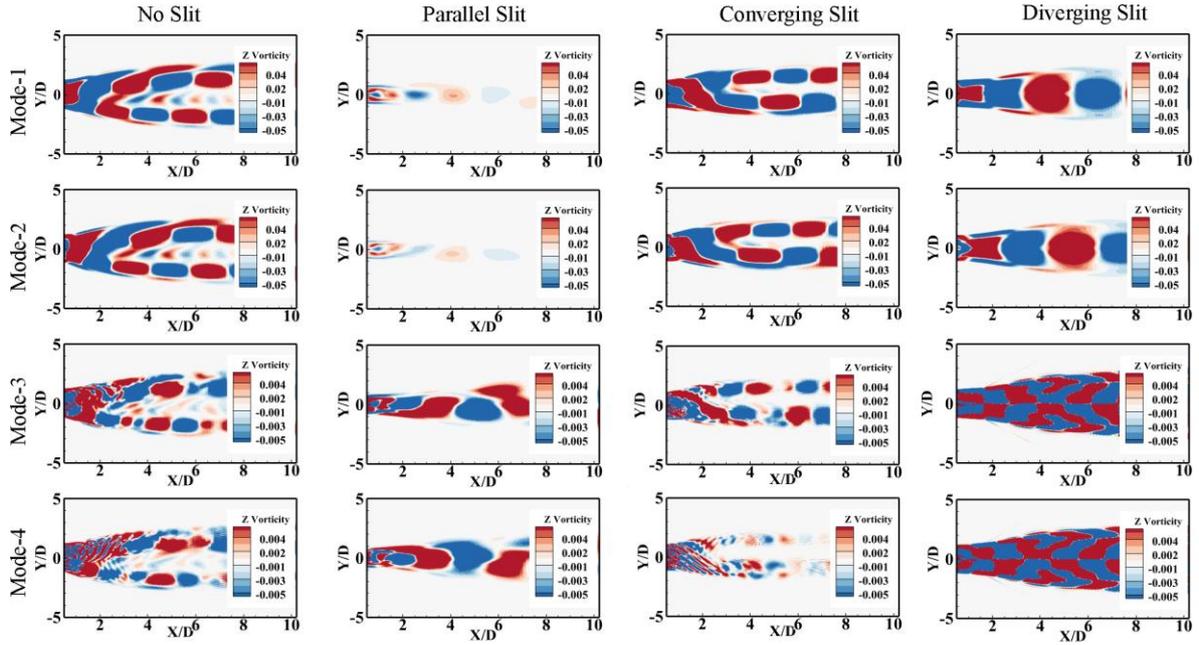

FIG. 13. First four most energetic modes for energy based Proper Orthogonal Decomposition (POD) of the flow field behind the cylinder undergoing transverse vibrations. [Contours represent the component corresponding z vorticity]

Figure 13 depicts the first four most-energy-containing modes for different slits. For the cylinder with no slit case, 90% of the fluctuating energy is distributed among the first 6 POD modes. Among those six modes, the first two modes are the most energy-containing modes. The third and the fourth mode are dominant in the near wake region and diffuses almost symmetrically downstream. For Diverging slit, the first two modes contain ~90% of the fluctuation energy with a relatively higher decay rate of eigenmode energy and represent the symmetric structure about the centerline of the wake, resembling the Von-Karman vortex street with alternate shedding. For the converging slit, more than 50% energy is contained in the first mode and resembles the second mode (with slightly lower energy). The third and fourth mode diffuses downstream away from the cylinder.

The parallel slit shows a different energy distribution among the POD modes. The cylinder with a parallel slit shows the lowest decay rate of the eigenmode energy and takes 16 POD modes to complete 93% of the total energy. Here, the energy is equally distributed in the first four modes, as shown in Fig. 12 (c). Then again, the next four POD modes contain similar energy (lower than the first four modes). This can be attributed to the closely spaced vortex shedding for the case of the parallel slit (as can be seen in the instantaneous vorticity contour plots in Fig. 10) and represents the contribution of the shear layer over the vortex shedding. Mishra et al.[27] reported that such reduction in the energy of the modes is due to the interaction between the primary vortex (through the cylinder) and secondary vortex (through the slit) and the irregular shedding.

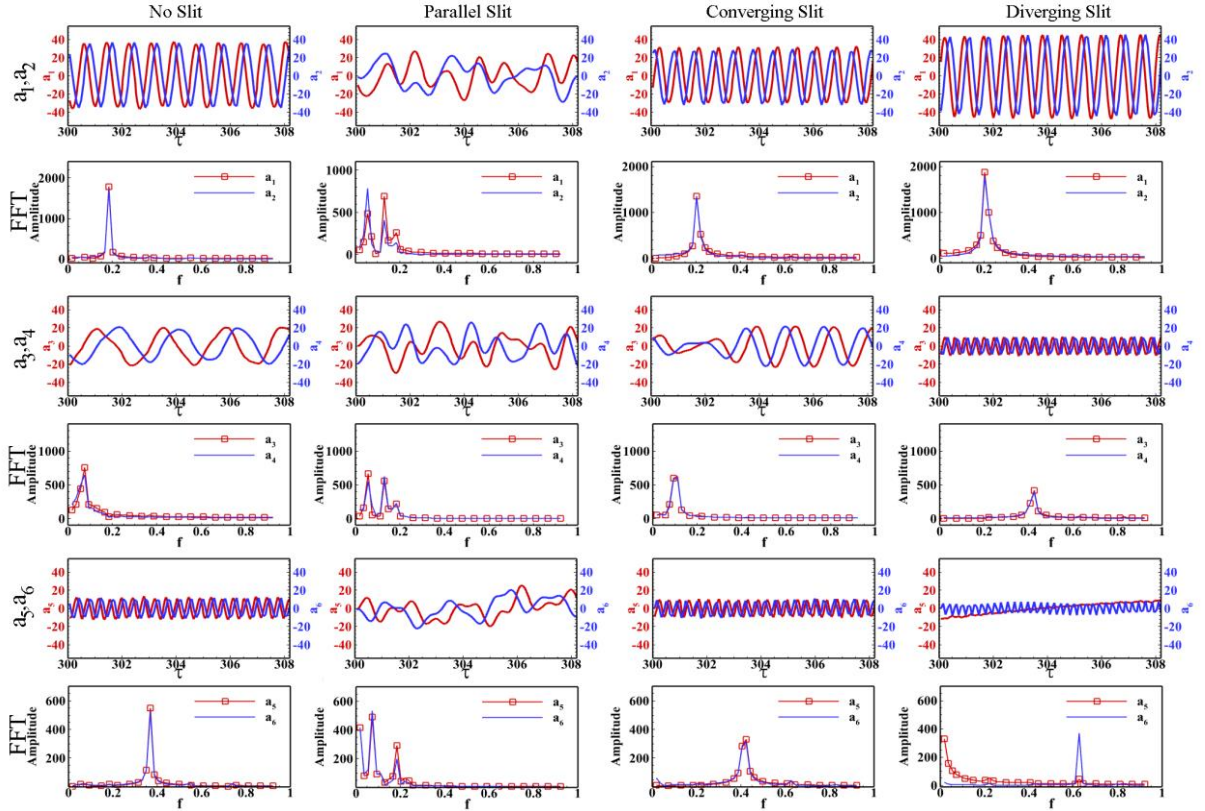

FIG. 14. Variation of the POD coefficients and their corresponding FFT for different cases [Red line represents 'Odd numbered coefficients' and Blue line represents 'Even numbered coefficients']

Figure 14 reports the variation of the POD coefficient and their corresponding peaks of the dominant frequencies as tabulated in Table IV. The alternate POD modes are coupled to their successive modes with the same dominant frequency and are lagged by a phase difference. These coupled pairs contain a similar amount of energy observed from the energy distribution plot in Fig. 12 (c). For the parallel slit, similar to the stationary cases, different peaks of the dominant frequencies appear. The FFT spectrum loses its characteristics of super-harmonics (which are easily identifiable in the case of convergent, divergent, and no-slit cases). The first four modes exhibit similar frequency peaks and amounts of energy which are much lesser than the other cases, while the 5$^{th}$ and 6$^{th}$ modes represent the contribution of the structures associated with even lesser frequencies. Although small in frequency, these structures also contain a significant amount of fluctuation kinetic energy compared to the first four flow modes (refer to Table IV for the individual energy contributions).

TABLE IV. Peaks of Dominant Frequency for POD coefficients

|  | $a_1$ | $a_2$ | $a_3$ | $a_4$ | $a_5$ | $a_6$ |
|---|---|---|---|---|---|---|
| **No Slit** | 0.185 | 0.185 | 0.071 | 0.071 | 0.356 | 0.356 |
| **Parallel Slit** | 0.057<br>0.128<br>0.171 | 0.057<br>0.128<br>0.171 | 0.057<br>0.128<br>0.171 | 0.057<br>0.128<br>0.171 | 0.014<br>0.071<br>0.157<br>0.185 | 0.014<br>0.071<br>0.157<br>0.185 |
| **Converging Slit** | 0.200 | 0.200 | 0.100 | 0.100 | 0.413 | 0.413 |
| **Diverging Slit** | 0.214 | 0.214 | 0.413 | 0.413 | 0.014 | 0.613 |

Thus, in the purview of the above discussions, it can be concluded that the parallel slit distributes the energy of the flow behind the wake into the number of structures associated with the lower frequencies and shows the lowest decay rate of energy. It results in a more stabilized wake and a significant reduction in the aerodynamic loading over the cylinder leading to suppressed VIV. Further, in the coming sections, the observations mentioned above are confirmed through the parametric variation in the slits, which includes the variation in the slit area-ratio (for Converging and divergent slit), slit-width (for parallel slit), and the slit angle (for the parallel slit).

## B. Effect of Slit Area Ratio

To confirm the parallel slit performance over the converging and diverging slit further, the slit-area ratio (as defined in the results section) is varied to 4 different values, i.e. 1.5, 2.3, 3.0, and 4.0 (for converging and diverging slits), and compared the results with the parallel slit (with slit area ratio 1.0) for slit width, s/D = 0.16 in Fig. 15. At lower area ratios (say Area ratio of 1.5), the mass flow from the slit is almost the same as a parallel slit. Hence, both the converging slit and the diverging slit contribute in the same way to suppress VIV. But, as the area ratio increases, the exit area for the converging slit gets reduced, and the effectiveness of the slit reduces, resulting in the higher lift coefficient values and their corresponding higher amplitude oscillations. Similar is valid for the diverging slit, where the entry area of the slit gets reduced, and less fluid flows through the slit. It results in a higher lift and reduced VIV suppression. For all the area-ratios, the cylinder with the parallel slit is associated with the lower aerodynamic lift coefficient values and works efficiently in suppressing the transverse oscillations of the slit cylinder.

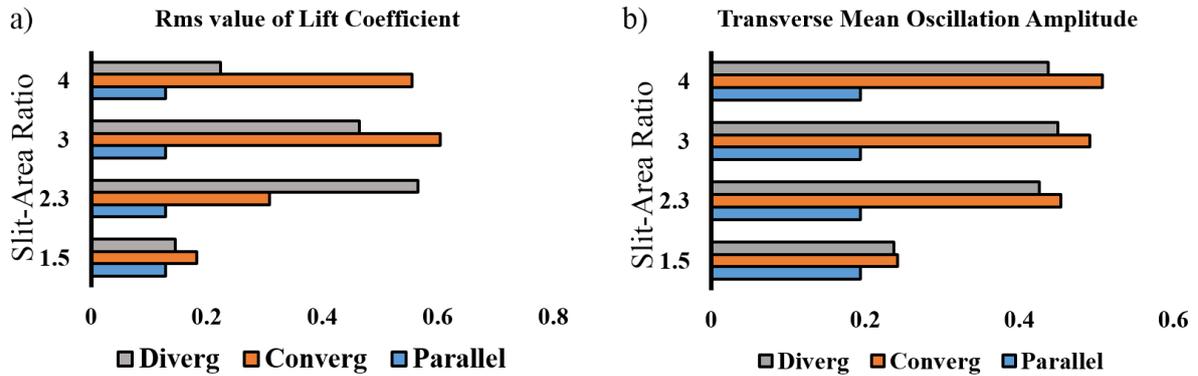

FIG. 15. Effect of slit-area ratio on (a) Aerodynamic lift coefficient, (b) Transverse mean oscillation amplitude

## C. Effect of Slit Width

As the parallel slit performs better than the converging slit, and the suppression gets more assertive with the increase in Re, we have assessed the effect of slit width (for parallel slit) on the suppression of transverse oscillation at Re = 500. The simulation extends to four different slit widths ranging from s/D = 0.12 to s/D = 0.24 with a step of 0.04, and the results are depicted in Fig. 16. At the lower slit widths (s/D < 0.16), the flow from the slit adds an extra amount of flow energy to the main flow, which results in the reduction of the oscillation amplitude (as compared to the no slit case). But, it doesn't affect the vortex shedding much, and the cylinder oscillates with the S+P pattern of vortex shedding similar to the no-slit case (as can be seen in Fig. 16 (b)). With the increase in the slit width, the amplitude of $C_L$ also decreases and starts to lose its periodic nature, which can be attributed to the increased interactions between the primary vortices (from the cylinder) with the secondary vortices (from the slit). During this, the cylinder oscillations are in phase with the lift force (refer to the phase portraits in Fig. 16(c)), and the vortex shedding is observed to be chaotic and no longer retains S+P shedding. Although, due to the reduction in the lift coefficient, the cylinder oscillates with the lower transverse amplitude. This reduction in $C_L$ occurs till s/D = 0.20. At s/D = 0.20, a $90^o$ phase shift is observed between the cylinder oscillations and its lift force and the lowest transverse oscillation amplitude. For s/D > 0.20, the amplitude of $C_L$ and Y/D starts to increase again. The observations agree well with Baek et al.[21] in their study of the effect of slit width with different mass ratios.

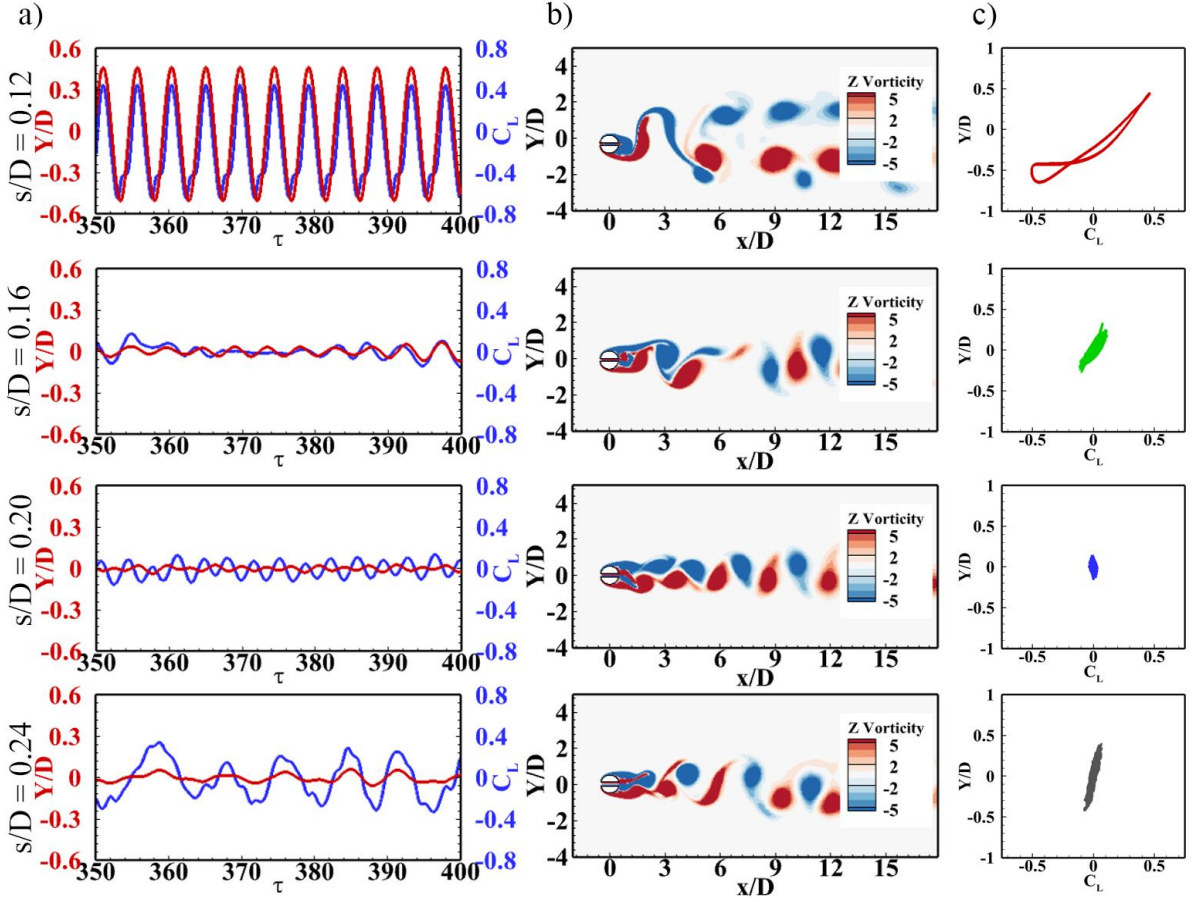

FIG. 16. a) Temporal distribution of the lift coefficient and their corresponding transverse amplitude for the modified cylinders with different slit widths, b) Instantaneous vorticity contour, and c) Phase Portraits [at Re = 500, s/D = 0.16, $U_r = 4.44$ and $m^* = 2$]

## D. Effect of Slit Angle

Finally, the slit angle is varied from $0^o$ (corresponding to the parallel slit) to $30^o$ with a step of $5^o$, and the flow characteristics are investigated at Reynolds no. of 500 with a slit width s/D = 0.16. Figures 17 (a) and (b) show the vortex shedding and the corresponding Reynolds shear stress distribution behind the slit cylinder at different slit angles. Noticeably, the chaotic shedding (at $\theta = 0^o$) gradually changes to form a Von-Karman street type of shedding with the increase in the slit angle (till $\theta = 20^o$). For the parallel slit ($\theta = 0^o$), the Reynolds shear stress is anti-symmetrically distributed along the wake axis. The increase in the slit angle makes the Reynolds shear stress distribution asymmetrical along the axis, and the peaks of the lift coefficients start to rise till $\theta = 5^o$. Further increase of angle ($\theta > 5^o$) results in the decrease of the lift coefficient, and the lower peak value of the Reynolds shear stress in the wake denoted the wake-stabilization. This reduced lift coefficient can be attributed to the establishment of the periodic alternate Von-Karman Street like shedding. The variation of $C_L$ becomes periodic and results in periodic transverse oscillation [$5^o \geq \theta \geq 20^o$]. This matches well with the similar observations by Mishra et al.[27] in their study of the stationary slit cylinder. At the slit angle of $\theta = 20°$, the cylinder oscillations are observed to be out of phase with the lift force as shown in Fig. 17 (c). At this slit angle, the cylinder oscillates with the minimum transverse oscillation amplitude. With a further increase in slit angle [$\theta > 20^o$], there is a considerable reduction in the slit flow, causing the lower pressure recovery and the higher global instability (as shown in Fig. 18 (a) and (b)) behind the cylinder. The vortex shedding mode changes from 2S to 2P, where two pairs of the counter-rotating vortices are shed in one cycle of oscillation, and there is a considerable increase in the amplitude of the lift coefficient and the transverse amplitude (see Fig. 18 (c)).

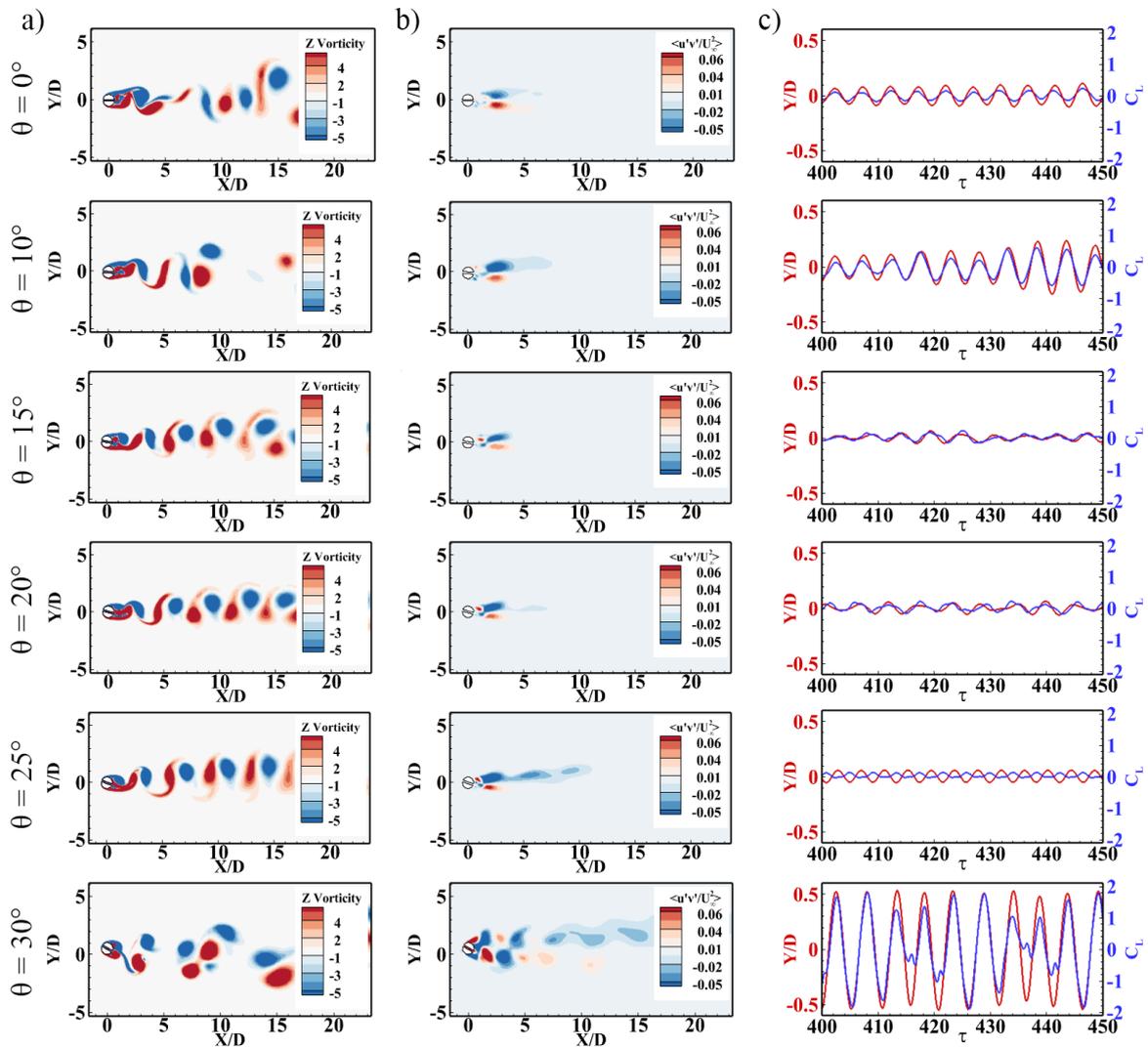

FIG. 17. Flow field behind the oscillating cylinder with a parallel slit at different slit angles: (a) Instantaneous vorticity contour, (b) Distribution of the Reynolds shear stress in the wake, and (c) Temporal evolution of the aerodynamic lift coefficient and the corresponding normalized transverse amplitude of the oscillation [at Re = 500, s/D = 0.16, $U_r$ = 4.44 and $m^* = 2$]

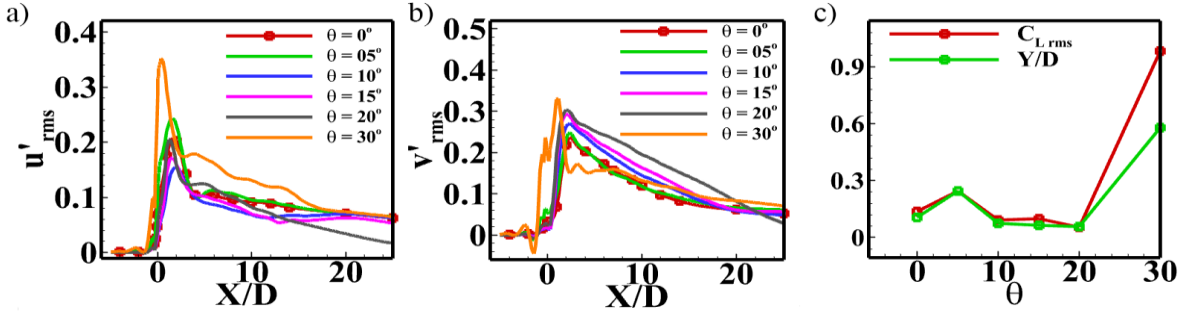

FIG. 18. The maximum rms velocity fluctuations along the line of constant x/D for different slits: (a) Axial velocity fluctuations (u'), (b) Transverse velocity fluctuations (v'); and (c) variation of the rms value of the mean coefficient and the mean transverse oscillation amplitude with different slit angles [at Re = 500, s/D = 0.16, $U_r = 4.44$ and $m^* = 2$]

To summarize, this study assesses the flow field associated with the elastically mounted slit cylinder at a Reynolds number of 500. First, the simulations are performed on the stationary cylinder with different slits to develop a basic understanding of the underlying physics. The simulations are then performed for the transversely oscillating cylinder. The flow features have been assessed using computational techniques such as FFT spectrum, flow visualization techniques, and the decomposition of the flow field into several POD modes based on their fluctuation-kinetic energy contribution. All these methods shed light on the various aspects of the flow around the slit cylinder and suggest that the parallel slit heavily modifies the flow behind the oscillating cylinder due to the strong interaction between the slit flow and the primary flow over the cylinder. Further, the parametric variation study is performed to accommodate the various parametric effects.

## V. Conclusion

The present study investigates the effectiveness of the different slit shapes (i.e. converging slit, diverging slit, and the parallel slit) in suppressing the VIV of an elastically mounted cylinder, bounded to oscillate in the transverse direction only. The stationary cases with slit cylinders show the shift and elongation of the primary vortices downstream the cylinder, resulting in pressure recovery on the cylinder base, a stabilized wake, and a significant reduction in aerodynamic loading on the cylinder. Further, the observations are verified by performing the moving slit cylinder simulations. The cylinder with the no-slit oscillates and sweeps the P+S pattern of the vortex shedding, while the additional flow from the parallel slit changes it to 2S mode for the parallel slit. The converging slit cylinder exhibits no change in the vortex shedding and retains the S+P pattern, while for the diverging slit, the Karman-Vortex street is formed (2S pattern). The cylinder with the parallel slit oscillates with a much lower amplitude than the rest and exhibits the minimum global instability of the cylinder wake. With the increase in the Reynolds number, the VIV suppression effectiveness also increases. POD reveals that the parallel slit cylinder retains the lowest decay rate of the eigenmodes energy. Through the observation and findings, the current study concludes that the parallel slit performs better than the converging or diverging slit case. Moreover, the effectiveness of the converging and diverging slit reduces heavily with the increase in slit-area ratio due to less flow passing from the slit. Also, initially, the slit cylinder retains the alternate shedding, which weakens further with the increase of slit-angle, but after a specific value, the shedding switches from 2S mode to 2P mode, and the oscillation amplitude increases rapidly. The detailed discussion finally explains and portrays the wake vortex dynamics for different slits.

## Data Availability

The data that support the findings of this study are available from the corresponding author upon reasonable request.

## Acknowledgments

The authors would like to acknowledge National Supercomputing Mission (NSM) for providing computing resources of 'PARAM Sanganak' at IIT Kanpur, which is implemented by C-DAC and supported by the Ministry of Electronics and Information Technology (MeitY) and Department of Science and Technology (DST), Government of India. We would also like to acknowledge the IIT-K Computer Centre (www.iitk.ac.in/cc) for providing the resources to perform the computation work. This support is gratefully acknowledged.

**Appendix**

We take Grid-C as the base grid and approximate the error in Grid-D compared to Grid-C from the Richardson error estimator, defined by,

$$E_1^{fine} = \frac{\varepsilon_{DC}}{1 - r_{DC}^o} \qquad (A1)$$

Error in Grid-B, compared to the solution of Grid-C, is approximated by coarse-grid Richardson error estimator, which is defined as,

$$E_2^{coarse} = \frac{r^o \varepsilon_{CB}}{1 - r_{CB}^o} \qquad (A2)$$

Where r is the grid refinement ratio between the two consecutive grids defined as,

$$r_{i+1,i} = \frac{h_{i+1}}{h_i}$$

The error ($\varepsilon$) is estimated from the solutions of two consecutive grids by,

$$\varepsilon_{i+1,i} = \frac{f_{i+1} - f_i}{f_i} \qquad (A3)$$

Grid-Convergence Index (GCI), which accounts for the uncertainty in the Richardson error estimator, is calculated for fine-grid and coarse-grid as given[48-49],

$$GCI_{fine} = F_s |E_1^{fine}| \tag{A4}$$

$$GCI_{coarse} = F_s |E_2^{coarse}| \tag{A5}$$

Before using Richardson extrapolation for GCI, we have also checked for the convergence conditions as follows:

- Monotonic convergence: 0 < R < 1,
- Oscillatory convergence: R < 0,
- Divergence: R > 1

Here, R, the convergence ratio, is defined as,

$$R = \frac{\varepsilon_{i,i-1}}{\varepsilon_{i+1,i}} \tag{A6}$$

Further to calculate the order of accuracy, $L_2$ norm of the errors between is grids is calculated as,

$$L_2 = \sqrt{\left(\sum_{i=1}^{N} |\varepsilon_{i+1,i}|^2 / N\right)} \tag{A7}$$

Where N is the total number of grid points considered for the calculation of $L_2$ norm. Figure 4 (c) depicts the $L_2$ the norm against the grid spacing h on a log-log scale. Further, the slope of $L_2$ norm is compared with the theoretical slope of order 2. The order of accuracy (or the slope of $L_2$ norm) is found to be 1.927. This value is used for GCI calculations of different grids. Also, the value of R confirms the monotonic nature of convergence for the grid, as can be seen in Table II.

# References

[1]C. H. K. Williamson, and A. Roshko, "Vortex formation in the wake of an oscillating cylinder," J. Fluids Struct. **2** (4), 355-381(1988). DOI: 10.1016/S0889-9746(88)90058-8

[2]M. Liu, W. Yang, W. Chen, H. Xiao, and H. Li, "Experimental investigation on vortex-induced vibration mitigation of stay cables in long-span bridges equipped with damped crossties," J. Aero. Eng. **32**(5), 04019072 (2019). DOI: 10.1061/(ASCE)AS.1943-5525.0001061

[3]M. M. Bernitsas, K. Raghavan, Y. Ben-Simon, and E. M. H. Garcia, "VIVACE (Vortex Induced Vibration Aquatic Clean Energy): A New Concept in Generation of Clean and Renewable Energy from Fluid Flow," ASME. J. Offshore Mech. Arct. Eng. **130**(4), 041101 (2008). DOI: 10.1115/1.2957913

[4]T. Zhou, S. M. Razali, Z. Hao, and L. Cheng, "On the study of vortex-induced vibration of a cylinder with helical strakes," J. Fluids Struct. **27**(7), 903-917 (2011). DOI: 10.1016/j.jfluidstructs.2011.04.014

[5]S. K. Kumar, C. Bose, S. F. Ali, S. Sarkar, and S. Gupta, "Investigations on a vortex induced vibration based energy harvester," App. Phys. Letters **111**(24), 243903 (2017). DOI: 10.1063/1.5001863

[6]Y. Yu, F. Xie, H. Yan, Y. Constantinides, O. Oakley, and G. E. Karniadakis, "Suppression of vortex-induced vibrations by fairings: a numerical study," J. Fluids Struct. **54**, pp.679-700 (2015). DOI: 10.1016/j.jfluidstructs.2015.01.007

[7]D. Kumar, M. Mittal, and S. Sen, "Modification of response and suppression of vortex-shedding in vortex-induced vibrations of an elliptic cylinder," Int. J. Heat and Fluid Flow **71**, 406-419 (2018). DOI: 10.1016/j.ijheatfluidflow.2018.05.006

[8]W. L. Chen, D. B. Xin, F. Xu, H. Li, J. P. Ou, and H. Hu, "Suppression of vortex-induced vibration of a circular cylinder using suction-based flow control," J. Fluids Struct. **42**, pp.25-39 (2013). DOI: 10.1016/j.jfluidstructs.2013.05.009

[9]H. Zhu, T. Tang, H. Zhao, and Y. Gao, "Control of vortex-induced vibration of a circular cylinder using a pair of air jets at low Reynolds number," Phys. Fluids **31**(4), 043603 (2019). DOI: 10.1063/1.5092851

[10]O. Cetiner, and D. Rockwell, "Streamwise oscillations of a cylinder in a steady current. Part 1. Locked-on states of vortex formation and loading," J. Fluids Mech. **427**, pp.1-28 (2001). DOI: 10.1017/S0022112000002214

[11]S. Choi, H. Choi, and S. Kang, "Characteristics of flow over a rotationally oscillating cylinder at low Reynolds number," Phys. Fluids **14**, pp. 2767-2777 (2002). DOI: 10.1063/1.1491251

[12]H. Zhu, and J. Yao, "Numerical evaluation of passive control of VIV by small control rods," App. Ocean Res. **51**, pp.93-116 (2015). DOI: 10.1016/j.apor.2015.03.003

[13]J. Y. Hwang, K. S. Yang, and S. H. Sun, "Reduction of flow-induced forces on a circular cylinder using a detached splitter plate," Phys. Fluids **15**, pp.2433-2436 (2003). DOI: 10.1063/1.1583733

[14]Y. Z. Law, and R. K. Jaiman, "Staggered Grooves for the Suppression of Vortex-Induced Vibration in Flexible Cylinders," In Int. Conf. Offshore Mech. and Arctic Eng. **58776**, p. V002T08A027. ASME (2019). DOI: 10.1115/OMAE2019-95649

[15]T. Zhou, S. F. Razali, Z. Mohd Hao, L. Cheng, "On the study of vortex-induced vibration of a cylinder with helical strakes," J. Fluids Struct. **27**, pp. 903-917 (2011). DOI: 10.1016/j.jfluidstructs.2011.04.014

[16]T. Igarashi, "Flow Characteristics around a Circular Cylinder with a Slit: 1st Report, Flow Control and Flow Patterns," Bulletin of JSME **21** (154), 656-664 (1978). DOI: 10.1299/jsme1958.21.656

[17]T. Igarashi, "Flow characteristics around a circular cylinder with a slit: 2nd report, effect of boundary layer suction," Bulletin of JSME **25**(207), 1389-1397 (1982). DOI: 10.1299/jsme1958.25.1389

[18]C. O. Popiel, D. I. Robinson, and J. T. Turner, "Vortex shedding from a circular cylinder with a slit and concave rear surface," Appl. Sci. Research **51** (1), 209-215 (1993). DOI: 10.1007/BF01082539

[19]J. F. Olsen, and S. Rajagopalan, "Vortex shedding behind modified circular cylinders," J. Wind Eng. and Indus. Aerodyn., **86** (1), 55-63 (2000). DOI: 10.1016/S0167-6105(00)00003-9

[20]S. Dong, G. S. Triantafyllou, and G. E. Karniadakis, "Elimination of vortex streets in bluff-body flows," Phys. Review Letters **100** (20), 204501 (2008). DOI: 10.1103/PhysRevLett.100.204501

[21]H. Baek, and G. E. Karniadakis, "Suppressing vortex-induced vibrations via passive means," J. Fluids Struct. **25** (5), 848-866 (2009). DOI: 10.1016/j.jfluidstructs.2009.02.006

[22]W. Junwei, Y. Jinwen, H. Yi, and B. Feng, "Experimental study of slit cylinder vortex shedding in circulating water channel," In Proceedings of 2012 Int. Conf. on Measur. Info. and Control, IEEE, 225-229 (2012). DOI: 10.1109/MIC.2012.6273318

[23]D. L. Gao, W. L. Chen, H. Li, and H. Hu, "Flow around a slotted circular cylinder at various angles of attack," Exp. Fluids **58**(10), 1-15 (2017). DOI: 10.1007/s00348-017-2417-8

[24]D. L. Gao, W. L. Chen, H. Li, and H. Hu, "Flow around a circular cylinder with slit," Exp. Therm. Fluid Sci. **82**, 287-301 (2017). DOI: 10.1016/j.expthermflusci.2016.11.025

[25]A. Mishra, M. Hanzla, and A. De, "Passive control of the onset of vortex shedding in flow past a circular cylinder using slit," Phys. Fluids **32** (1), 013602 (2020). DOI: 10.1063/1.5132799

[26]B. Sharma, and R. N. Barman, "Steady laminar flow past a slotted circular cylinder," Phys. Fluids **32**(7), 073605 (2020). DOI: 10.1063/5.0007958

[27]A. Mishra, and A. De, "Suppression of vortex shedding using a slit through the circular cylinder at low Reynolds number," Euro. J. Mech.-B/Fluids **89**, 349-366 (2021). DOI: 10.1016/j.euromechflu.2021.06.009

[28]I Orlanski, "A simple boundary condition for unbounded hyperbolic flows," J. Comp. Phys. **21**(3), 251-269 (1976). DOI: 10.1016/0021-9991(76)90023-1

[29]G. Biswas, and H. Chattopadhyay, "Heat transfer in a channel with built-in wing-type vortex generators," Int. J. Heat Mass Transfer **35**(4), 803-814 (1992). DOI: 10.1016/0017-9310(92)90248-Q

[30]H. G. Weller, G. Tabor, H. Jasak, and C. Fureby, "A tensorial approach to computational continuum mechanics using object-oriented techniques," Comp. Phys. **12**(6), 620-631 (1998). DOI: 10.1063/1.168744

[31]C. Kassiotis, "Which strategy to move the mesh in the Computational Fluid Dynamic code OpenFOAM," Report École Normale Supérieure de Cachan, 1–14 (2008). URL: http://perso.crans.org/kassiotis/openfoam/movingmesh.pdf

[32]H. Jasak, "Dynamic Mesh Handling in OpenFOAM," 47th AIAA Aero. Sci. Meeting including The New Horizons Forum and Aero. Expos. **341** (2009). DOI: 10.2514/6.2009-341

[33]D. D. Kosambi, Statistics in function space, J. Indian Math. Soc. **7**, 76-88 (1943). URL: http://repository.ias.ac.in/99240/1/Statistics_in_function_space.pdf

[34]J. L. Lumley, "The structure of inhomogeneous turbulent flows," Atmos. Turb. and Radio Wave Prop. **790**, 166-178 (1967). URL: https://ci.nii.ac.jp/naid/10012381873/

[35]L. Sirovich, "Turbulence and the dynamics of coherent structures. I. Coherent structures," Quarterly of Appl. Math. **45**(3), 561-571 (1987). DOI: https://www.ams.org/journals/qam/1987-45-03/S0033-569X-1987-0910462-6/S0033-569X-1987-0910462-6.pdf

[36]B. Bhatia and A. De, "Numerical Study of Trailing and Leading Vortex Dynamics in a Forced Jet with Coflow," Comp. Fluids **181**, 314-344 (2019). DOI: 10.1016/j.compfluid.2019.02.001

[37]R. Soni and A. De, "Role of Jet spacing and Strut geometry on the formation of large scale structures and mixing characteristics," Phys. Fluids **30**(5), 056103 (2018). DOI: 10.1063/1.5026375

[38]R. Soni and A. De, "Investigation of Mixing Characteristics in Strut Injectors Using Model Decomposition," Phys. Fluids **30**(1), 016108 (2018). DOI: 10.1063/1.5006132

[39]R. K. Soni, N. Arya, and A. De, "Characterization of turbulent supersonic flow over a backward facing step through POD," AIAA J. **55**(5), 1511-1529 (2017). DOI: 10.2514/1.J054709

[40]P. Das and A. De, "Numerical study of flow physics in supersonic base-flow with mass bleed," Aero. Sci. Tech. **58**, 1-17 (2016). DOI: 10.1016/j.ast.2016.07.016

[41]P. Das and A. De, "Numerical investigation of flow structures around a cylindrical afterbody under supersonic condition," Aero. Sci. Tech. **47**, 195-209 (2015). DOI: 10.1016/j.ast.2015.09.032

[42]G. Kumar, A. De, and H. Gopalan, "Investigation of flow structures in a turbulent separating flow using Hybrid RANS-LES model," Int. J. Num. Meth. for Heat and Fluid Flow **27**(7), 1430-1450 (2017). DOI: 10.1108/HFF-03-2016-0134

[43]H. T. Ahn, and Y. Kallinderis, "Strongly coupled flow/structure interactions with a geometrically conservative ALE scheme on general hybrid meshes," J. Comp. Phys. **219**(2), 671-696 (2006). DOI: 10.1016/j.jcp.2006.04.011

[44]I. Borazjani, and F. Sotiropoulos, "Vortex-induced vibrations of two cylinders in tandem arrangement in the proximity–wake interference region," J. Fluids Mech. **621**, 321–364 (2009). DOI: 10.1017/S0022112008004850

[45]Y. Bao, C. Huang, D. Zhou, J. Tu, and Z. Han, "Two-degree-of-freedom flow-induced vibrations on isolated and tandem cylinders with varying natural frequency ratios," J. Fluids Struct. **35**, 50-75 (2012). DOI: 10.1016/j.jfluidstructs.2012.08.002

[46]A. K. Soti, J. Zhao, M. C. Thompson, J. Sheridan, and R. Bhardwaj, "Damping effects on vortex-induced vibration of a circular cylinder and implications for power extraction," J. Fluids Struct. **81**, 289-308 (2018). DOI: 10.1016/j.jfluidstructs.2018.04.013

[47]CFD, ICEM, "ANSYS ICEM CFD User's Manual," (2016).

[48]P. J. Roache, "Perspective: A method for uniform reporting of grid refinement studies," J. Fluids Eng. **116**, 405–413 (1994). DOI: 10.1115/1.2910291

[49]P. J. Roache, "Quantification of uncertainty in computational fluid dynamics," Annu. Rev. Fluid Mech. **29**(1), 123–160 (1997). DOI: 10.1146/annurev.fluid.29.1.123

[50]M. S. M. Ali, C. J. Doolan, and V. Wheatley, "Grid convergence study for a two-dimensional simulation of flow around a square cylinder at a low Reynolds number," In Seventh Int. Conf. on CFD in The Minerals and Process Indus. (ed. PJ Witt & MP Schwarz), 1-6 (2009). URL: https://www.cfd.com.au/cfd_conf09/PDFs/136ALI.pdf

[51]R. Govardhan, and C. H. K. Williamson, "Modes of vortex formation and frequency response of a freely vibrating cylinder," J. Fluid Mech. **420**, 85-130 (2000). DOI: 10.1017/S0022112000001233

[52]Y. Delaunay, and L. Kaiktsis, "Control of circular cylinder wakes using base mass transpiration," Phys. Fluids **13**(11), 3285-3302 (2001). DOI: 10.1063/1.1409968

[53]M. Nishioka, and H. Sato, "Mechanism of determination of the shedding frequency of vortices behind a cylinder at low Reynolds numbers," J. Fluid Mech. **89**(1), 49-60 (1978). DOI: 10.1017/S0022112078002451